\documentclass[reprint, prl, superscriptaddress, floatfix]{revtex4-2}

\usepackage{hyperref}
\hypersetup{
    colorlinks=true,
    linkcolor=blue,
    citecolor=red,      
    urlcolor=blue,
    pdftitle={An open-source robust machine learning platform for real-time detection and classification of 2D material flakes},
    pdfpagemode=FullScreen,
    }
\usepackage{graphicx}
\usepackage{amsmath}
\usepackage{amsfonts}
\usepackage{color}
\usepackage{verbatim}
\usepackage{upgreek}
\usepackage{bm}
\usepackage{textcomp}
\usepackage[symbol*]{footmisc}
\usepackage{float}
\usepackage{xcolor}
\usepackage{multirow}

\DefineFNsymbolsTM{otherfnsymbols}{
  \textdagger    \dagger
  \textdaggerdbl \ddagger
  \textsection   \mathsection
  \textparagraph \mathparagraph
 }
\setfnsymbol{otherfnsymbols}

\newcommand{\beq}{\begin{equation}}
\newcommand{\eeq}{\end{equation}}
\newcommand{\bea}{\begin{eqnarray}}
\newcommand{\eea}{\end{eqnarray}}
\newcommand{\bal}{\begin{align}}
\newcommand{\eal}{\end{align}}

\newcommand{\fig}[1]{Fig.\,\ref{#1}}

\newcommand{\WSe}{WSe$_2$}

\begin{document}
\title{An open-source robust machine learning platform for real-time detection and classification of 2D material flakes}

\author{Jan-Lucas Uslu} 
\email{jan-lucas.uslu@rwth-aachen.de}
\author{Taoufiq Ouaj} 
\author{David Tebbe}
\affiliation{2nd Institute of Physics and JARA-FIT, RWTH Aachen University, 52074 Aachen, Germany}
\author{Alexey Nekrasov}
\affiliation{Visual Computing Institute,
RWTH Aachen University, 52074 Aachen, Germany}
\author{Jo Henri Bertram}
\author{Marc Sch\"utte}
\affiliation{2nd Institute of Physics and JARA-FIT, RWTH Aachen University, 52074 Aachen, Germany}
\author{Kenji Watanabe}
\affiliation{Research Center for Electronic and Optical Materials, National Institute for Materials Science, 1-1 Namiki, Tsukuba 305-0044, Japan}
\author{Takashi Taniguchi}
\affiliation{Research Center for Materials Nanoarchitectonics, National Institute for Materials Science,  1-1 Namiki, Tsukuba 305-0044, Japan}
\author{Bernd Beschoten} 
\affiliation{2nd Institute of Physics and JARA-FIT, RWTH Aachen University, 52074 Aachen, Germany}
\affiliation{JARA-FIT Institute for Quantum Information, Forschungszentrum J\"ulich GmbH and RWTH Aachen University, 52074 Aachen, Germany}
\author{Lutz Waldecker} 
\email{waldecker@physik.rwth-aachen.de}
\affiliation{2nd Institute of Physics and JARA-FIT, RWTH Aachen University, 52074 Aachen, Germany}
\author{Christoph Stampfer}
\affiliation{2nd Institute of Physics and JARA-FIT, RWTH Aachen University, 52074 Aachen, Germany}
\affiliation{Peter Gr\"unberg Institute (PGI-9) Forschungszentrum J\"ulich, 52425 J\"ulich, Germany}

\begin{abstract}

The most widely used method for obtaining high-quality two-dimensional materials is through mechanical exfoliation of bulk crystals.
Manual identification of suitable flakes from the resulting random distribution of crystal thicknesses and sizes on a substrate is a time-consuming, tedious task.
Here, we present a platform for fully automated scanning, detection, and classification of two-dimensional materials, the source code of which we make openly available.
Our platform is designed to be accurate, reliable, fast, and versatile in integrating new materials, making it suitable for everyday laboratory work.
The implementation allows fully  automated scanning and analysis of wafers with an average inference time of 100 ms for images of 2.3 Mpixels. 
The developed detection algorithm is based on a combination of the flakes' optical contrast toward the substrate and their geometric shape.
We demonstrate that it is able to detect the majority of exfoliated flakes of various materials, with an average recall (AR50) between 67\% and 89\%. 
We also show that the algorithm can be trained with as few as five flakes of a given material, which we demonstrate for the examples of few-layer graphene, WSe$_2$,  MoSe$_2$, CrI$_3$, 1T-TaS$_2$ and hexagonal BN. 
Our platform has been tested over a two-year period, during which more than $10^6$ images of multiple different materials were acquired by over 30 individual researchers.  
 
\end{abstract}


\maketitle
\date{\today}

\section{Introduction}

Two-dimensional (2D) materials offer an unprecedented opportunity to artificially engineer van der Waals (vdW) heterostructures for a wide range of applications, as well as to study fundamental material properties \cite{Geim2013}. 
These vdW heterostructures can range from monolayers encapsulated in hexagonal boron nitride (hBN) for improved electronic and optical properties and spatial homogeneity~\cite{Dean2010, Neumann2015,Banszerus2015Jul, Cadiz2017, Raja2019}, to stacks containing up to ten layers, such as various semiconductors, graphitic gates, and hBN dielectrics and spacer layers \cite{Xu2020, Park2021, Liu2021}.
Despite progress in the growth of 2D materials \cite{Backes2020Jan,Burton2023Jan} and their subsequent fabrication of vdW heterostructures on a wafer scale \cite{Liu2021_review, Xu2022}, the exfoliation and mechanical stacking of 2D crystals remains the method of choice for basic research and proof-of-principle devices \cite{Liu2016}.

Currently, most researchers manually identify suitable flakes by scanning the exfoliated crystals on the substrate under a microscope \cite{Novoselov2005_2dcrystal}, which is a time-consuming and tedious task.
This is particularly true for air-sensitive materials, which need to be exfoliated and identified in an inert environment. 
Therefore, automating the identification of flakes has the potential to drastically improve the efficiency of sample preparation.

In general, the automatic detection of flakes can be achieved by recording optical microscope images with a motorized microscope and a digital camera and subsequent image analysis by a computer algorithm \cite{Ryu2022}. 
Several groups have reported demonstrations of such detection algorithms, e.g. using classical machine-learning tools such as support vector machines or K-means \cite{Masubuchi2018, Lin2018Dec, Li2019Sep}.
These methods rely on the fact that 2D materials exist in integer numbers of layers and their optical contrast values with respect to a substrate are therefore discrete \cite{Blake2007}.
Other groups have employed neural networks to achieve the task of flake detection \cite{Saito2019Dec, Han2020Jul}. 
This approach provides greater versatility, as the algorithm can learn to identify flakes by different types of features and established networks can be extended to the detection of new materials. 
However, neural networks tend to need large amounts of labeled data, a requirement which is impractical, especially when dealing with materials with a low yield of mono to few-layer flakes upon exfoliation.
Training the detection algorithm on a small number of manually identified flakes is therefore highly desirable for the integration of new materials on a reasonable timescale as well as for the implementation of the algorithm on different optical setups.
For neural networks, this has been attempted using transfer learning \cite{Masubuchi2020Mar, Han2020Jul}.
However, the challenge of requiring large quantities of labeled data can also be alleviated by directly using inductive biases based on physical features, such as the optical contrast of the flakes, to reliably classify instances.

\begin{figure*}[!bth]
    \begin{center}       \includegraphics{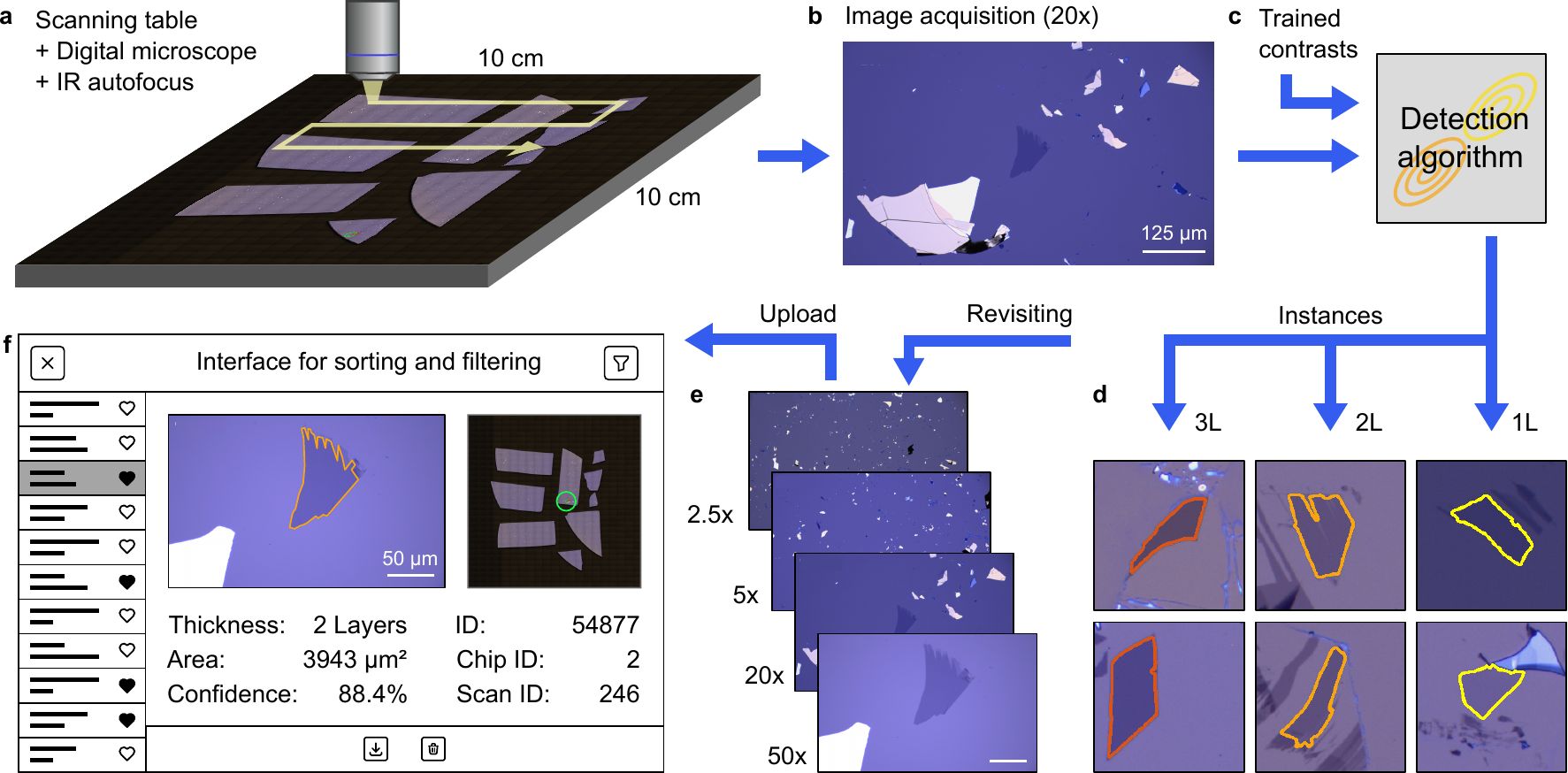}
        \caption{Workflow overview for the example of few-layer graphene. (a) Si/SiO$_2$ wafer pieces with exfoliated graphite are placed on a microscope stage. An infrared (IR) autofocus module allows to keep the images in all magnifications in focus in realtime. The full range of the scanning table is scanned in the lowest magnification (2.5$\times$) from which the positions of the wafer pieces are inferred. (b) Images are then recorded in higher magnification (here 20$\times$). (c) These images are analyzed by the detection algorithm using their optical contrast (see section II and \fig{fig:detector}). (d) The detection algorithm classifies flakes by their optical contrast and assigns the layer number (1L, 2L, etc.) together with an instance mask (colored outlines). The sidelength of the boxes are 75 \textmu m. (e) After the scan has finished, images of each detected flake are taken in all possible magnifications.  (f) All images as well as their metadata (i.e. area, number of layers etc.) are uploaded to an interactive website for inspection and selection by the user. An example website can be accessed at \cite{website}.}
        \label{fig:pipeline}
    \end{center}
\end{figure*}

Next to an efficient detection algorithm, an automated flake search system needs to be fast, reliable, and compatible with the workflow of the other fabrication steps in order to fully replace manual search.
In this work, we present an automated workflow for the scanning, detection, and classification of exfoliated 2D materials on $\mathrm{SiO_2}$ using classical machine learning models. 
We address the full task of flake detection starting with the exfoliation of crystal flakes on optimized substrates, over real-time detection and classification, to the storage of detected flakes in a database.
The main focus is on the development of a robust and fast pipeline suitable for daily use in the laboratory as a truly competitive alternative to manual searching by a researcher; an overview of the process workflow is given in section I. 
At the center of the implementation is the detection algorithm, which utilizes a Gaussian mixture model (GMM) to fit clusters in the optical contrast space, as well as a simple logistic classifier to assign a confidence value based on the geometric features of the detected flakes.
The details of the implementation are given in section II, while the code is made publicly available at \cite{code}.
The training of our algorithm is performed with approximately five example images of a given material and number of layers, as described in section III. 
While the contrast values of many 2D materials, including few-layer graphene, WSe$_2$, MoSe$_2$, CrI$_3$, 1T-TaS$_2$ as well as hBN form discrete clusters and can therefore naturally be implemented with our method, practically continuous distributions, such as those of hBN between 10-40 nm thickness, can also be fitted. 
The biggest challenge to the stability of the detection algorithm is variations in the oxide thickness of the substrates, which typically vary between wafer batches from different growth runs, but can be compensated for in the training process.  
To quantify the performance of the detection algorithm and to allow a fair comparison to other implementations, we present metrics of the detection in section IV.  
We identify the average recall (AR50) as particularly relevant to the problem, as it describes the fraction of exfoliated flakes which have been properly detected.
It varies between materials and flake sizes from 63\% for small \WSe\ up to 100\% for flakes larger than 400~\textmu m$^2$, demonstrating that large flakes are practically always detected by our algorithm.

\section{I. Workflow overview}

A comprehensive overview of the workflow is presented in \fig{fig:pipeline}.
We start with mechanical exfoliation of 2D crystals using tape onto Si/SiO$_2$ substrates \cite{Huang2015}, resulting in flakes of arbitrary size and shape.
By choosing a particular oxide thickness, thin-film interference leads to a large optical contrast in the visible range \cite{Blake2007, Muller2015}.
The choice of oxide thickness depends on the material and will be discussed in section III.
The wafer pieces are placed on the motorized scanning table (M\"arzh\"auser SCAN 100$\times$100) of the microscope setup (Nikon Eclipse).
The microscope is equipped with a motorized revolver holding five microscope objectives between 2.5$\times$ and 100$\times$ magnification, a digital color camera (The Imaging Source DFK 33UX174), and an infrared autofocus module (Nikon LV-DAF), which works independently of the camera to stabilize the focal plane.
The entire setup is placed in a glovebox with an inert nitrogen atmosphere. 

Our program starts by scanning the entire area of the scanning table in the smallest magnification installed (CFI T Plan EPI 2.5$\times$), see \fig{fig:pipeline}a. 
The recorded images are downsampled and stitched together to create an overview image of all wafer pieces on the scanning table. 
The downsampling is not strictly necessary but reduces the filesize of the resulting overview image. 
From the overview image, the position and orientation of the wafer pieces is deduced by thresholding the image using Otsu's method \cite{Otsu1979}, yielding a binary mask of wafer and background.
The overview mask is later used to define the positions which will be scanned in higher magnification for the detection of the flakes, and thus allows using wafer pieces of arbitrary shape while minimizing the overall scan time.

For the following flake search, we use a magnification of 20$\times$ (CFI TU Plan Fluor EPI 20$\times$), which is a trade-off between the time needed and the minimal size of detectable flakes (\fig{fig:pipeline}b).
The entire surface of all wafer pieces is then scanned, and each image is analyzed at run-time. 
We detect flakes of interest by applying an image detection algorithm  (\fig{fig:pipeline}c) based on the combination of a Gaussian mixture model (GMM) and a shape discriminator, which is the main focus of this paper and will be described in detail in section II. 

For each detected flake  (\fig{fig:pipeline}d), the image, together with an instance mask, metadata, and the global position, are saved, and a marker is placed at its position in the overview image.
Images not containing detected flakes are discarded.

At the end of the scan, every detected flake is revisited in all magnifications (2.5$\times$, 5$\times$, 20$\times$, 50$\times$, 100$\times$) to record images with them in the center of the field of view, see \fig{fig:pipeline}e.
Images recorded in larger magnifications help for later validation of the quality of the detected flakes by the user, e.g. the cleanliness of the surface, whereas images in smaller magnifications are essential in order to locate the flakes for further processing, i.e. for stacking of flakes into vdW heterostructures.

The collection of images, positions, and metadata, such as their thickness, size, or aspect ratio, are uploaded to a database on a server; see \fig{fig:pipeline} f. 
The data is accessible via a web interface, which also allows for filtering and sorting.
The interface enables for a fast and efficient selection of flakes for a particular device structure.
It also encourages sharing of materials between different users or projects, as often different material thicknesses or shapes are needed (e.g. few-layer graphene for back gates and contacts, mono- or bilayer graphene for heterostructure). 
Thus, easy access to the gathered data is an integral part of a productive workflow; a demo version can be accessed at \cite{website}.

\section{II. Detection algorithm}

\begin{figure*}[!bth]
    \begin{center}
        \includegraphics{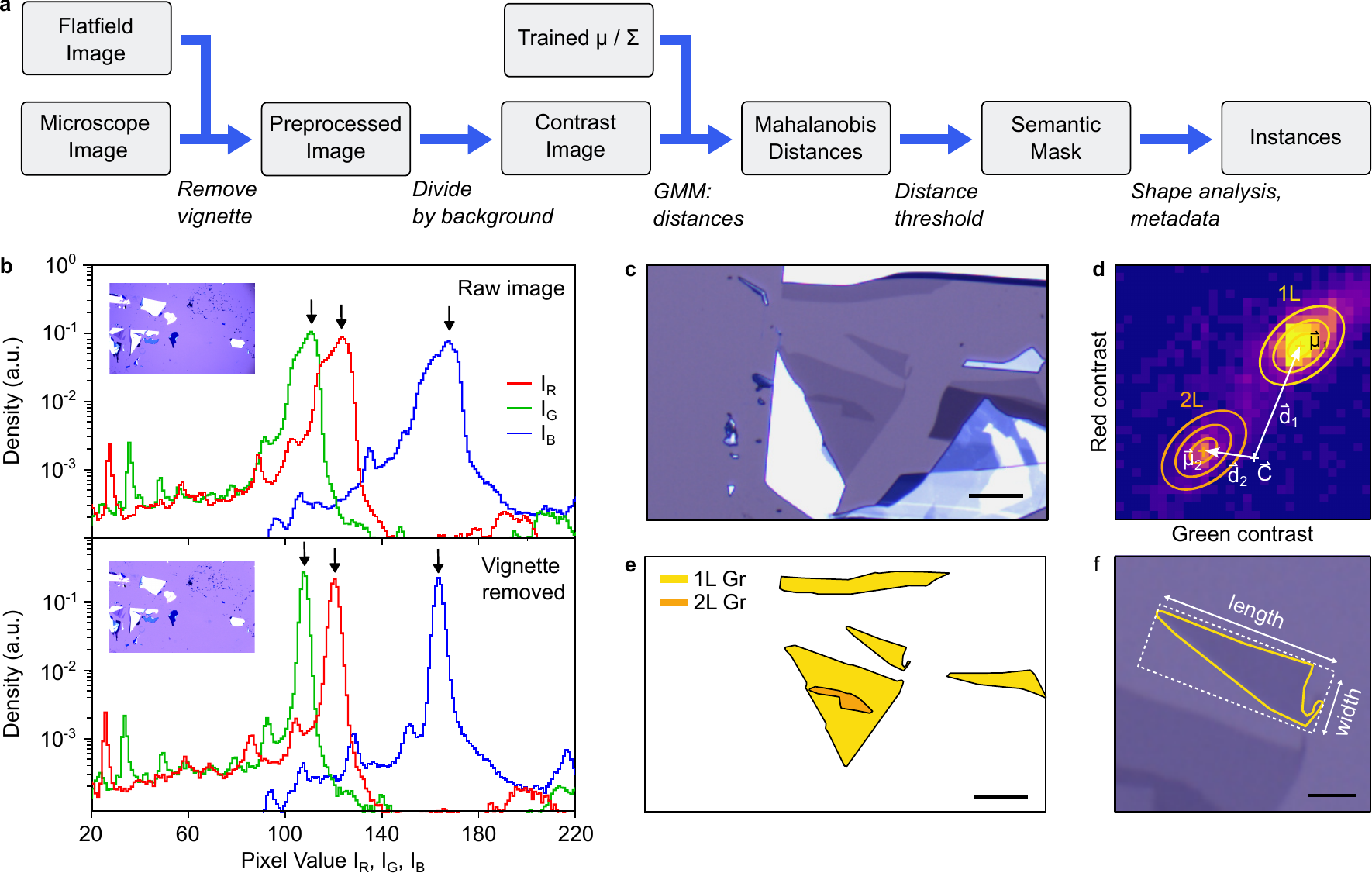}
        \caption{
        Detection Algorithm. a) Workflow of the detection algorithm (see text for details). b) Color histograms of a raw image (top) and the preprocessed image (bottom), in which the vignette has been removed. The main features (arrows) originate from the bare Si/SiO$_2$ wafer piece and therefore correspond to the background color. The respective images are shown in the inset; the colors have been exaggerated for better visibility of the effect. c) Image containing different graphene flakes (scale bar: 20 \textmu m). d) Sketch of the working principle of the Gaussian mixture model for layer identification. The contrast distributions of each layer are given as means $\vec{\mu_i}$ (crosses) and covariance matrices $\Sigma$ (ellipses). A measured color contrast of a pixel $\vec{C}$ has a distance $\vec{d_i}$ to the mean $\vec{\mu_i}$ of each Gaussian (measured in standard deviations). e) Final segmentation mask of the image shown in c. f) Extraction of metadata from detected instances, including bounding box and side lengths. Scale bar: 10 \textmu m. 
        }
        \label{fig:detector}
    \end{center}
\end{figure*}

The detection algorithm introduced here is based on the extraction of two general and robust features of exfoliated flakes of 2D materials. 
First, the optical contrast of the flakes towards their substrate~\cite{Blake2007, Gao2008, Li2019Sep} is calculated.
We then compare the contrast of individual pixels to pre-determined average contrasts of each number of layers, which were extracted beforehand using the GMM. 
This enables effective segmentation and classification of various areas within the image.
Although the optical contrast discriminates different flake thicknesses, it is not sufficient for a reliable flake detection, as other objects, such as shadows and tape residues, can have a similar contrast and might thus falsely be classified as flakes.
This is most notable when searching for materials with  small contrasts, such as graphene or, in particular, few-layer hBN. 
Therefore, in s second step, we employ a logistic classifier, which is based on the geometric shape of the detected flakes.
This additional classifier assigns a false positive probability to each flake previously detected by the GMM.

A flowchart of all individual computational steps of the detection algorithm is shown in~\fig{fig:detector}a.
The detection algorithm starts with the image taken by the microscope during the scan in higher magnification (here 20$\times$).
Three corrections are applied to the image.
First, the dark counts of the camera are subtracted from each RGB channel.
Second, we account for the effect of vignetting, that is, an inhomogeneous illumination of the image with decreased brightness toward the edges (compare the images in the inset of \fig{fig:detector}b).
This is corrected by dividing the image by a flatfield image (pixel-wise) and multiplying it by the mean pixel value of the respective color channel, obtaining a preprocessed image $\vec{I}_{\mathrm{p}} (x,y)$ where $\vec{I} (x,y) = (I_\mathrm{R}, I_\mathrm{G}, I_\mathrm{B})(x,y)$ is the pixel value for each RGB color channel at position $(x,y)$~\cite{Seibert1998}.
The flatfield image is taken once for each substrate, containing only the bare Si/SiO$_2$ substrate.
To demonstrate the effect of the vignette correction, the color histograms of an example image before and after its removal are shown  in~\fig{fig:detector}b.
Each color channel exhibits a prominent mode (see black arrows) which is at least two orders of magnitude stronger than others.
These  correspond to the color of the background, i.e. the intensity of the bare Si/SiO$_2$ substrate, which takes up the largest area of the image.
Note that saturated areas of the image, e.g. from bulk crystals, can take up large areas, too, but are excluded by only using color values within the range of 20 to 230.
The effect of the image correction can be clearly seen in the reduction of the widths of the histograms.
After removal of the vignette, a global background color $\vec{I}_{\mathrm{bg}}$ is extracted as the mode of the distribution, indicated by the arrows in~\fig{fig:detector}b.
Third, we apply a median blurring (5$\times$5 kernel) to reduce camera noise while preserving edges to a good degree.
An example is shown in \fig{fig:detector}c.

\begin{figure*}[!tbh]
    \begin{center}
        \includegraphics[width=1.0\textwidth]{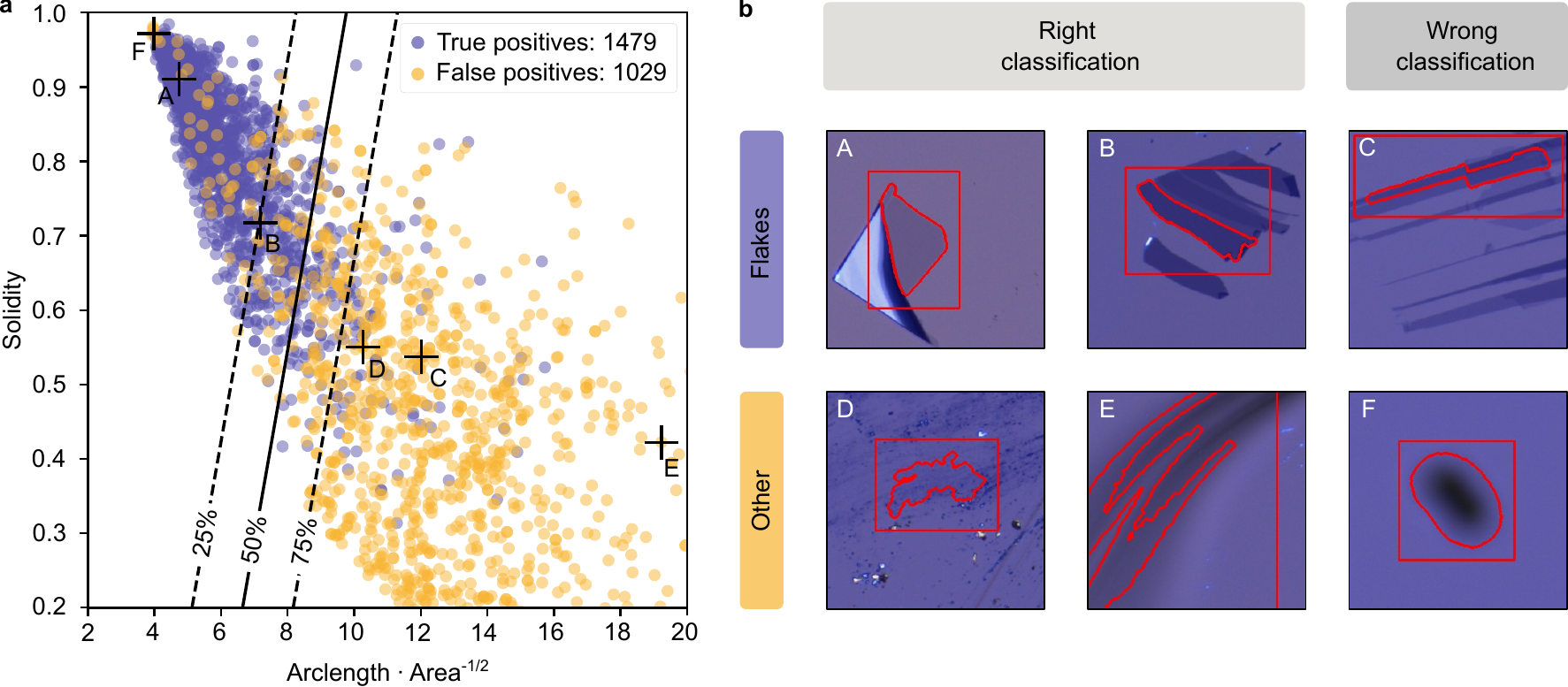}
        \caption{Shape analysis and assignment of false positive probability. a) Scatter plot of solidity vs arclength$\cdot$area$^{-1/2}$ of the masks of 2508 detected flakes. Real flakes (blue) and others (orange), such as shadows or tape residue, have been annotated by hand and show clearly different distributions, which can be used to assign a false positive likelihood. The solid and dashed black line gives the 25, 50 and 75$\%$ probability boundaries. b) Examples of six detected instances and their classification using the logistic regression (see text for details). Their positions in panel a) is indicated by capital letters. The sidelength of the boxes is 115 \textmu m. 
        }
        \label{fig:FPdet}
    \end{center}
\end{figure*}

From the pre-processed images, the contrast $\vec{C}$ of each pixel with coordinates $(x,y)$ is calculated:
\begin{equation}
    \Vec{C}(x,y) = (\vec{I}_{\mathrm{p}}(x,y) - \vec{I}_{\mathrm{bg}}) \oslash \vec{I}_{\mathrm{bg}},
    \label{eq:contrast}
\end{equation}
where $\oslash$ denotes component-wise division.
Using the contrast instead of raw intensities makes the detection process independent of lighting conditions.
The resulting contrast images can now be segmented by the trained contrast values (see section III for details on the training). 
The trained contrasts are defined by a set of Gaussians $S$ with means $\Vec{\mu}$ and covariance matrices $\Sigma$, which describe the variation of contrast values for each number of layers $n_L$ (see sketch in \fig{fig:detector}c).
The likelihood of a pixel belonging to a certain layer is then proportional to its Mahalanobis distance (i.e. the distance in standard deviations) to the mean values $\Vec{\mu}$.
For each pixel, the Mahalanobis distance $d_S$ to each set $S$ is calculated by:
\begin{equation}
    d_{S}(\Vec{C}, \Vec{\mu}, \Sigma) = \sqrt{(\Vec{C}-\Vec{\mu})\Sigma^{-1}(\Vec{C}-\Vec{\mu})}.
\end{equation}
The layer number with the smallest Mahalanobis distance is then assigned to each respective pixel in the segmented image.

Pixels, for which none of the distances are within a certain range, $d_S > d_{\mathrm{max}}$ for all $S$, are classified as background. 
$d_{\mathrm{max}}$ can be adjusted to detect more or fewer pixels, but is typically taken as five standard deviations.

The resulting segmentation mask contains all detected flakes, but also wrongly assigned pixels due to shadows, tape residue or single pixels from camera noise.
Single stray pixels are removed by applying morphological operations, i.e. through erosion followed by dilation.

In the next step, the semantic mask is converted to instances.
As flakes do not overlap, we are able to extract instances by finding the connected components on the semantic mask.
This is done by using the Spaghetti labeling algorithm from Ref.\cite{Bolelli} implemented in OpenCV. 
Each component is post-processed by finding the outer contour and redrawing the component, which fills remaining holes.
The resulting masks are instances of the contrast-based flake detection; an example image and its corresponding segmentation mask are shown in \fig{fig:detector}d and \fig{fig:detector}e.
Metadata are extracted for each component, which include the size, aspect ratio, maximum and minimum side-lengths as well as the mask itself (\fig{fig:detector}f). 
These metadata are saved together with all images in the database. 

To filter out detected instances that do not correspond to true flakes, we implement a discriminator, which analyzes the shape of the detected instances.
It is based on the observation that falsely assigned instances tend to have curved and irregular outlines (see \fig{fig:FPdet}b). 
We quantify this observation using the solidity, i.e. the ratio of the area of the mask to the area enclosed by the convex hull, and the length of the perimeter (arclength) divided by the square root of the area.

\fig{fig:FPdet}a shows a scatter plot of both properties of 2508 instances detected as graphene, which have been labeled manually as 'flakes' (blue) and 'others' (orange). 
A separation of true and false positives is observed, with false positives generally having lower solidity and a larger arc length per square root area. 
Therefore, a logistic regression can be used to separate the two classes by assigning a probability to each instance and interpreting this as a false positive probability.
We use an L2 logistic classifier to fit the data; the decision boundary (50\% false positive probability), as well as the 25\% and 75\% probability boundaries, are shown as solid and dashed lines in \fig{fig:FPdet}a. 
The false positive probability for each instance is saved in the metadata without deleting any instances and can be used to sort and filter out flakes. 

Note that the two geometric feature dimensions used here are not exclusive. 
An additional feature dimension which has been used is the Shannon entropy of the area \cite{Masubuchi2018}.
The entropy threshold for filtering, however, is less general, as it depends on parameters such as the lighting conditions, the white balance, or the gain of the camera. 
In contrast, our classifier is material and microscope agnostic, as no direct information about the image is used, and it is directly interpretable. 
More complex models, such as neural networks, might perform even better than our method; however, they tend to either overfit the data, require large computational overhead, or introduce unnecessary complexity (millions of parameters versus two parameters in our case).
The performance of the detection scheme for different materials will be discussed in section IV.

\section{III. Training of the detection algorithm}

The algorithm training process starts by manually identifying approximately 5 flakes for each number of layers of interest, with a size of at least 200 \textmu m$^2$ each, corresponding to 800 pixels for a magnification of 20$\times$.
These images are then annotated semi-automatically using the watershed algorithm (implemented in OpenCV).
At this point, it is not needed to define the number of layers of the flakes, as this is inferred later. 
The vignette of these images is removed, and the color contrast of each pixel within the flakes is calculated as described in Equation~\ref{eq:contrast}.

\begin{figure}[!tbh]
    \begin{center}
        \includegraphics{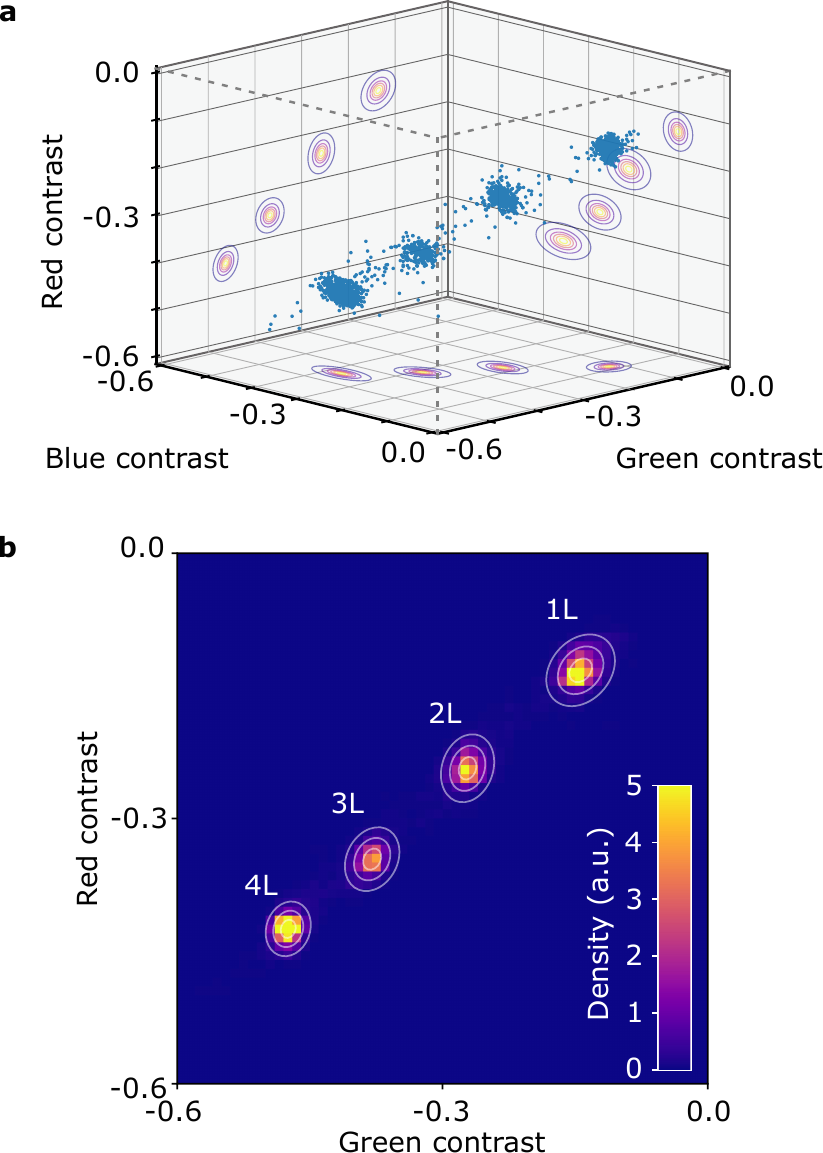}
        \caption{Training of the detection algorithm for the example of graphene on 90nm SiO$_2$/Si. a) The contrast values of few-layer graphene in RGB space show well-separated clusters. b) Corresponding color contrast density map in the red and green channel. The white lines depict the contours of the first, second and third standard deviation of a multi-dimensional Gaussian fit to the data. Four flakes of each number of layers have been used. }
        \label{fig:training}
    \end{center}
\end{figure}

The contrast distribution of one layer to four layers of graphene is shown in \fig{fig:training}a (four flakes each have been used).
The contrast values group into four distinct and well separated clusters, as described in previous publications \cite{Masubuchi2019, Li2019Sep, Sterbentz2021Mar}.
For easier comparability, we only show two of the three color channels as heat maps in the following. 
For the case of graphene on 90 nm SiO$_2$, the heat maps of the red and green channels are shown in \fig{fig:othermat}b (see the discussion on oxide thickness variations below for the choice of color axes).

The three-dimensional color contrasts are then fit with three-dimensional Gaussians.
After defining the number of clusters, the expectation–maximization (EM) algorithm \cite{Pedregosa2011} is run and returns a set $S$ of parameters $\{\pi,\vec{\mu}, \Sigma\}$.
While $\vec{\mu}$ and $\Sigma$ are the parameters defining the position and shape of the Gaussian, the $\pi$ parameter is a measure for the class weight. It counteracts class imbalance by automatically scaling the Gaussians according to the estimated weight of each class. After training the model, the weights $\pi$ are discarded since we are only interested in the shape of the Gaussians, as the shape encodes information about the contrast distribution of each class. This leads to a model that is balanced with regard to the class distribution.
To increase the robustness of the fit, more components of the GMM can be added to fit the background noise distribution (we typically use a ratio of about 2 to 1).
Finally, the layer numbers are assigned to the clusters, which is done by the order of their contrasts.
Given enough context (flakes of different thicknesses), this is usually straightforward, as in the example of few-layer graphene, but it could also be supported by simulations \cite{Muller2015}.  
The two-dimensional fits and labels are overlaid on the graphene training data in \fig{fig:training}.

\subsection{Implementation of different materials}
The training of our detection algorithm can be performed, as demonstrated above for few-layer graphene, with any 2D material, which has a sufficiently large color contrast on a particular substrate and with a sufficiently large contrast difference between individual layers. 

\begin{figure*}[!tbh]
    \begin{center}

        \includegraphics[width=1.0\textwidth]{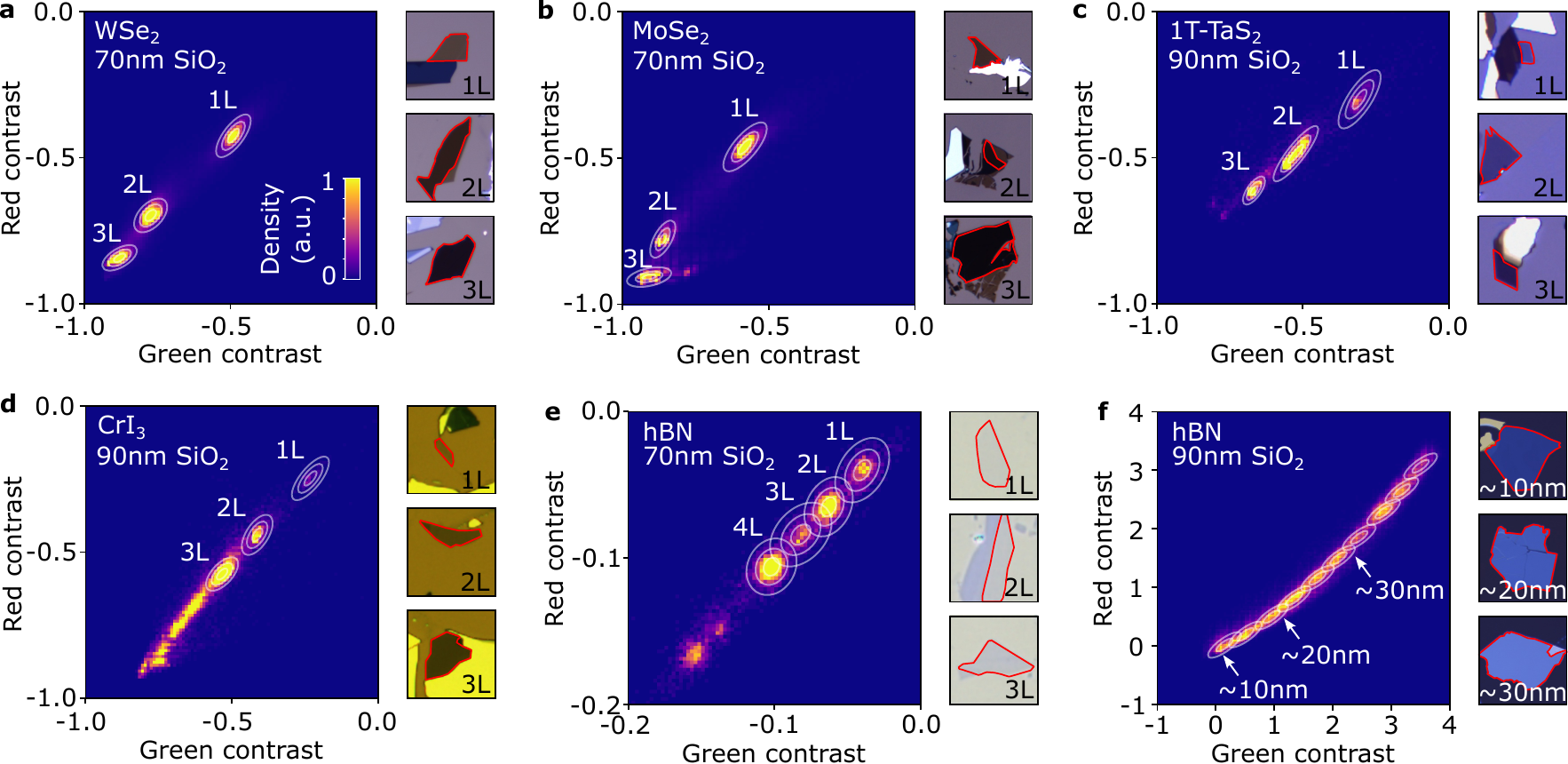}
        \caption{Heat maps of the color contrasts of four different 2D materials and their respective fits with the GMM (left panels) and example images of flakes of each material, which are used for the training (right panels). The side lengths of the boxes are 45~ \textmu m (a-e) and 90 \textmu m (f).
        (a-d) Well-separated clusters of the contrasts of few-layer WSe$_2$, MoSe$_2$, 1T-TaS$_2$ and CrI$_3$ are observed. (e) For thin flakes of the large band-gap insulator hBN, the contrast differences  of neighboring layers are small. In order to observe separated clusters, the images underlying the heatmap have been taken in higher magnification (50$\times$) and two exposures have been averaged to reduce camera noise. 
        (f) For thicker hBN flakes, the distribution becomes quasi-continuous in the range shown (up to $\approx40$nm). We fit this distribution with ten Gaussians, corresponding to intervals of approximately 4 nm.         
        }
        \label{fig:othermat}
    \end{center}
\end{figure*}

\fig{fig:othermat}a-e shows example images and contrast density maps (red and green contrast) obtained from exfoliated few-layer \WSe, MoSe$_2$, 1T-TaS$_2$, CrI$_3$ and few-layer hBN flakes on optimized Si/SiO$_2$ substrates. 
Note that for CrI$_3$, a 550 nm longpass filter was inserted into the microscope, as the material can degrade under blue-light illumination even in an inert nitrogen atmosphere.
For all five materials, the clustering of the contrasts is apparent, which makes it possible to employ a GMM to fit their distributions (see the white ellipses in the images). 
Note that for few-layer hBN, the absolute contrasts as well as the differences between neighboring layers are much smaller.
Even moderate noise in the images can therefore blur the differences between layers.
In order to observe clustering of the contrasts, which can be fitted with the GMM, we reduced the camera noise by averaging two consecutive exposures, and took images at 50$\times$ magnification, which decreases the influence of the flakes' edges. 
The contrast distribution, taken from three different flakes of hBN, is shown in \fig{fig:othermat}e.
The measured contrast values appear to have a small offset with respect to the background color, i.e. the first contrast step is comparatively bigger than all consecutive steps. 
This could be due to a small air gap between the flakes and the substrate. 
As we observe the offset to vary between flakes, the interpretation of the flake thickness needs to be done with care.

Lastly, we have applied the GMM to quasi-continuous contrast distributions, which appear if the contrast difference between adjacent layer numbers is well below the contrast deviation of a single layer.  
This is particularly relevant for the detection of thicker flakes of hBN, which are ubiquitously used to encapsulate and seal other 2D materials, as well as for gate dielectrics in 2D material devices.
In these circumstances, the exact thickness is usually irrelevant as long as it falls into a particular thickness range. 
\fig{fig:othermat}f shows the color contrast distribution of hBN flakes of thicknesses between few nm and approximately 40~nm. 
The distribution can be fit with the GMM, though more manual fine-tuning of the fit parameters, such as the number of components used and the type of covariance matrix, is needed.
Here, the distribution was fit with 10 Gaussians with the constraint of a single covariance matrix shared across all Gaussians, corresponding to thickness intervals of approximately 4 nm.

\begin{figure*}[!tbh]
    \begin{center}
        \includegraphics{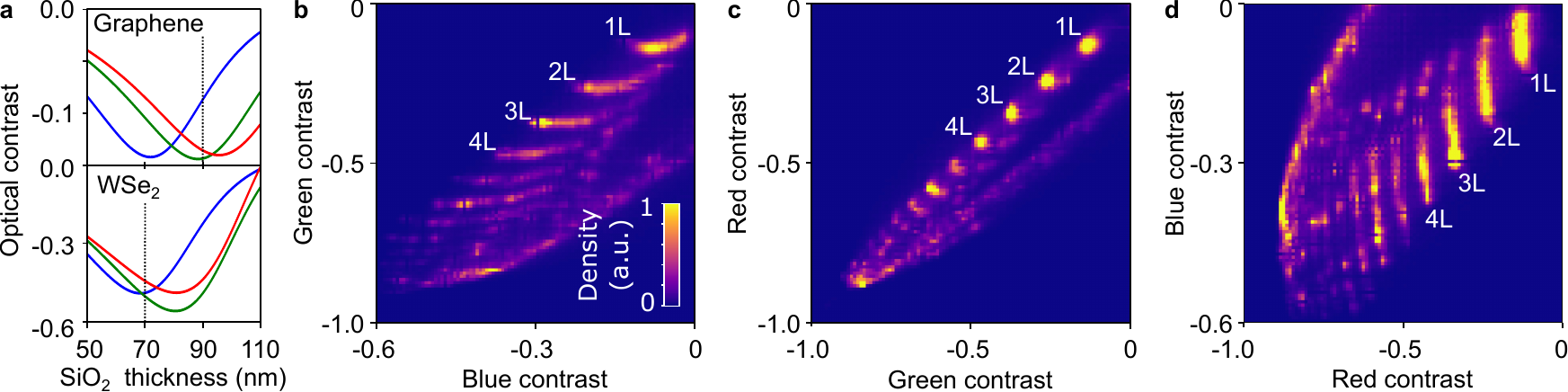}
        \caption{
        Influence of the silicon oxide thickness. a) Calculated optical contrast of monolayer graphene and \WSe\ on Si/SiO$_2$ in the RGB channels of a color camera. A high contrast value corresponds to good visibility and an easier detection. A large derivative (positive or negative) in a particular channel, however, renders the rule-based approach more unreliable in terms of thickness variations.  
        b)-d) The contrast distribution of few-layer graphene using 3600 Flakes from 11 different wafers of nominally 90 nm SiO$_2$ thickness. 
        The observed contrast distribution originates from variations in oxide thickness. 
        For graphene, it is largest in the blue channel, in agreement with the simulations shown in a). 
        }
        \label{fig:full_graphene}
    \end{center}
\end{figure*}

\subsection{Choice of Silicon oxide thickness}

The color contrast of 2D material flakes with respect to a Si/SiO$_2$ substrate arises from thin-film interference and therefore depends sensitively on the oxide thickness \cite{Gao2008, Li2019Sep}.
More specifically, the oxide thickness determines the spectral reflectivity $R(\lambda)$, where $\lambda$ denotes the wavelength, both of the substrate $R_{sub}$ and of the sample on the substrate $R_{sam}$. 
The color contrast  $\vec{C}$ furthermore depends on the spectral irradiance of the light source $I(\lambda)$ and the spectral sensitivity of the camera  $\vec{s}(\lambda)$ \cite{Muller2015}:
\begin{equation}
\begin{aligned}
    \vec{C} = & \left[ \left( \int d\lambda I(\lambda) \cdot R_{\mathrm{sam}}(\lambda, t_{ox}) \cdot \vec{s}(\lambda) \right) \right. \\
    & \left. \oslash  \left( \int d\lambda I(\lambda) \cdot R_{\mathrm{sub}}(\lambda, t_{ox}) \cdot \vec{s}(\lambda)  \right ) \right ]  - 1
\end{aligned}
\end{equation}

To simulate the color contrasts for different materials on Si/SiO$_2$ substrates, the reflectivities $R(\lambda)$ have been calculated using the transfer matrix method \cite{Alonso-Alvarez2018}.
Inputs to the calculation are the refractive indices of Si, SiO$_2$ and the 2D materials (here graphene and \WSe\ as examples) \cite{Weber2010, Green1995, Malitson1965, Li2018}, the camera sensitivity curve and an approximate spectrum of the illumination.

Figure \ref{fig:full_graphene}a shows the simulated color contrast of monolayer graphene and \WSe, respectively, on Si/SiO$_2$ wafers vs oxide thicknesses. 
The maximum contrast of each color channel is reached at a different oxide thickness and also depends on the material. 
Although a large contrast is a necessary requirement to be able to detect 2D material flakes, a large derivative of the contrast represents a challenge for the reliability of the contrast-based detection against variations of the oxide thickness. 
Such variations are common for the wafers produced  by dry oxidation in our facilities and for those of most commercial suppliers. 

Figures \ref{fig:full_graphene}b-d show heatmaps of the color contrast of few-layer graphene on (nominally) 90 nm oxide, obtained from 18 different exfoliations.
Instead of well-defined clusters, the color contrasts now vary along a characteristic curve.
The largest changes of contrast are observed along the blue axis, as expected from the simulation shown in \fig{fig:full_graphene}a.

For a stable operation of the contrast-based algorithm, we, therefore, choose an oxide thickness for which the color contrast varies as little as possible. 
For most materials, this condition cannot be fulfilled in all color channels simultaneously.
For graphene, we use wafers of 90~nm oxide thickness, for which at least two channels (red and green) are stable. 
For \WSe, the contrasts as well as their derivatives are more favorable at an oxide thickness of 70~nm. 
Of these two options, we find 90~nm oxide better suited for few layer 1T-TaS$_2$ and CrI$_3$, and 70 nm~oxide for the transition metal dichalcogneides (\WSe, MoSe$_2$) as well as few-layer hBN.

In addition, to make the algorithm more stable, some variations can on purpose be included in the training set by exfoliating onto different batches of wafers. 
While this works well to compensate for oxide variations of a few nanometers, wafers which are outside this range need to be sorted out, e.g. by performing ellipsometry, before use.

\section{IV. Evaluation}

In order to quantify the performance and establish a comparison between different approaches for the detection algorithm, meaningful metrics need to be used. 
Several recent publications used accuracy to compare the fidelity of the generated semantic masks \cite{Han2020Jul,Greplova2020Jun,Sterbentz2021Mar}.
This is not ideal, as the datasets exhibit a strong class imbalance between the background and the flakes of interest.
This imbalance can be on the order of $10^4$, i.e.  flakes only occupy less than 0.01\% of the image.
Using accuracy in this context will yield metrics in the realm of $99.99\%$, even if only a small fraction of flakes (or none) is classified as such.
To mitigate this, we evaluate our algorithm using metrics more suited for imbalanced datasets, in particular precision, recall and the Jaccard index (IoU).
In addition to these metrics, which are evaluated pixel-wise, we also use the instance-based average precision at 50\% (AP50) and average recall at 50\% (AR50) \cite{Lin2014}, which can be interpreted as the ratio of detected real flakes to false positives and the fraction of all real flakes detected by the algorithm. 
The latter two are determined using a modified MS COCO evaluator from the Detectron2 framework~\cite{wu2019detectron2}.

\begin{figure}[!t]
    \begin{center}
        \includegraphics{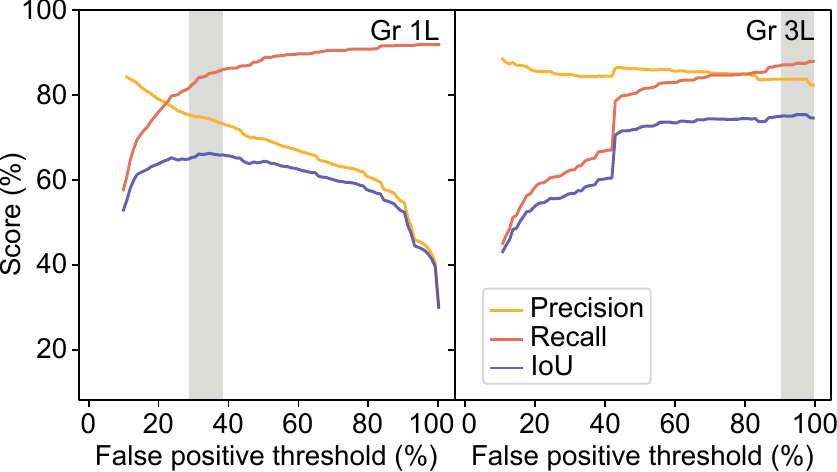}
        \caption{Metrics of the detection algorithm calculated for different levels of filtering based on the shape analysis. Flakes with a false-positive probability above the threshold are removed, which increases the precision (less false positives), but generally lowers the recall (found flakes). The gray shaded areas denote the level of false positive probability leading to the highest IoU for the two different numbers of layers. All scores have been evaluated pixel-wise.
        }
        \label{fig:metrics}
    \end{center}
\end{figure}

\subsection{Dataset}

We evaluated the algorithm on two materials, few-layer graphene and WSe$_2$
The few-layer graphene dataset consists of 1787 images collected over a period of one year from 12 different exfoliation runs, while the WSe$_2$ dataset contains 512 images collected from 14 different exfoliation runs over the same period.
Each image in the dataset is labeled with the number of layers present in the material, ranging from 1 to 4 layers for few-layer graphene and 1 to 3 layers for WSe$_2$. The images were collected using our automated setup, ensuring consistency, avoiding centering bias and have a resolution of 1920 $\times$ 1200 pixels.

The used images were collected by multiple researchers in our lab, resulting in varying conditions such as LED light aging, white balances, and substrate thickness. To ensure reliable training and testing of the models, we split the datasets into training and testing sets. Specifically, we used two exfoliation runs as the training set, and the rest of the runs were used for testing to better probe the generalization capabilities of the algorithm.
We did not employ any data augmentation techniques during the training process. 
The WSe$_2$ dataset comprises 92 training images and 420 test images, while the few-layer graphene dataset includes 425 training images and 1362 test images.

\subsection{Metrics}
The shape analysis introduced in section II assigns a false positive probability to each detected instance, which can be used for filtering out those above a certain threshold.  
As the value of this threshold affects the metrics of the algorithm, we first demonstrate its effect for different number of layers. 

\begin{table}
\begin{ruledtabular}
    \begin{tabular}{c cccc ccc}
       & \multicolumn{4}{c }{Few-layer graphene} & \multicolumn{3}{c}{WSe$_2$} \\
       & 1L & 2L & 3L & 4L & 1L & 2L & 3L \\  
       \hline
     Threshold  & 33       & 84     & 95      & 93     & 79   & 70   & 22 \\
     Precision  & 72.2     & 83.5   & 82.8    & 83.5   & 76.6 & 90.3 & 68.1\\
     Recall     & 83.8     & 83.3   & 86.9    & 87.5   & 59.9 & 60.5 & 61.2\\
     IoU        & 63.4     & 71.6   & 73.6    & 74.6   & 50.6 & 56.8 & 47.6\\
     \hline
     AP50  &  48.5  & 55.0 &  58.1  &  58.0   &   62.1  &  57.1  & 52.8 \\
     AR50  &  89.2 & 80.3  &  84.6  &  88.4   &   70.5  &  66.9  &  80.7   \\
    \end{tabular}
    \caption{Metrics of the detection algorithm. Top rows: the threshold at which the pixel-wise determined IoU is maximized and the respective precision, recall and IoU at this threshold for graphene (1-4 layers) and WSe$_2$ (1-3 layers). Bottom two rows: the instance-wise determined average precision (AP50) and average recall (AR50). All quantities are given in \%.
    \label{tab:metrics}
        }
\end{ruledtabular}
\end{table}

\begin{table}    
\begin{ruledtabular}
    \begin{tabular}{c c cccc}
        & size (\textmu m$^2$) & 50-100 & 100-200  & 200-400 & $>400$ \\
        \hline
        \multirow{2}{*}{AP50} 
        & Gr 1L  &  31.9 & 85.4 & 88.7 &  96.4  \\
        & WSe$_2$ 1L  &  52.7  &  78.5  &  77.8  &  100.0    \\
       \hline
       \multirow{2}{*}{AR50} 
        & Gr 1L  &  82.7  & 93.3 &  93.3  &  100.0    \\
        & WSe$_2$ 1L  &  62.3 & 81.9  &  81.0  &  100.0    \\
    \end{tabular}
    \caption{Size-dependence of the average precision (AP50) and the average recall (AR50). The detection becomes less reliable for smaller flakes as edges in the image make up for a larger fraction of the flake area. All quantities are in \%.   
    \label{tab:metricsII}
        }           
    \end{ruledtabular}
\end{table}

The precision, recall and IoU for the examples of monolayer and trilayer graphene are shown in \fig{fig:metrics} as a function of the threshold.
For both numbers of layers, the recall has its maximum for a threshold of 100\%, i.e. when no flakes are filtered at all.
This is expected, as the false positive (FP) filtering cannot increase the number of true positives. 
The precision, on the other hand, can increase as more FPs are filtered. 
The degree to which this happens markedly depends on the material: for monolayer graphene, which is optically thin, the precision increases sharply, whereas for trilayer graphene, only a small increase is observed. 
The reason is that contaminations tend to have a relatively low optical contrast and are therefore mostly classified as optically thin materials.
The filtering, therefore, has its main use case for monolayer and bilayer graphene as well as for other materials of low contrast, such as few-layer hBN. 
A point of good compromise can be defined by the maximum of the IoU.
We therefore provide the metrics at the threshold which maximizes the respective IoU in Table \ref{tab:metrics}. 

We stress that, from a practical standpoint, it is important that as many instances as possible are detected correctly, even if the detection mask is not perfect on a pixel level. 
We therefore calculate the instance-based precision and recall (AP50 and AR50).
The results are given for different materials and numbers of layers in Table \ref{tab:metrics}.

It is apparent that the metrics for all thicknesses of few-layer graphene are better than for \WSe, both in terms of IoU and AR50.
This is largely due to the significantly smaller average size of \WSe\ flakes: the detection mask and the labeled mask often differ on the very edges of the instances, leading to a lower precision and recall for smaller flakes.  
The size-dependence of the metrics is given in Table \ref{tab:metricsII}. 
The table shows that both, precision and recall, are significantly higher for flakes of sizes $>100$ \textmu m$^2$.
We note that we currently do not have enough labeled data for MoSe$_2$, TaS$_2$, and CrI$_3$ to report statistically significant metrics for those materials.
Yet, we expect the metrics to be very similar as their optical contrast distributions are comparable to graphene and WSe$_2$.
The reliable detection of few-layer hBN remains challenging due to its low optical contrast, which is often similar to the optical contrast of tape residue.

\subsection{Runtime}
Another important characteristic property of the algorithm is the computational time needed to process each image. 
The runtime was evaluated on the CPU (AMD Ryzen 5 3600) of a standard desktop computer. 
Averaged over 1000 images and searching for four different thicknesses, the runtime was 100 ms per image with resolution 1920 $\times$ 1200. 
This is significantly faster than other contrast-based implementations, which have reported inference times of about 3 seconds per image of size 1920$\times$1200~pixels \cite{Yang2020Sep}.
It is also faster than implementations based on neural networks, which have reported runtimes of similar magnitude (when run on a GPU) but were evaluated on images of 512$\times$512 and 224$\times$224~pixels, respectively \cite{Yang2022Aug,Han2020Jul}.
The short inference time means that the evaluation can be performed on-the-fly, i.e. during the time that the wafer pieces are scanned by the hardware.

Lastly, we discuss the time needed to fully complete a scan of wafer pieces to the final upload of images onto the server (see \fig{fig:pipeline}), which is the relevant time for a user.
The creation of the overview image takes approximately 9 minutes, as the entire area of the scanning table is scanned at low magnification. 
The scanning at higher magnification reaches a rate of approximately 2 images/s, corresponding to an area of about 25 cm$^2$/hour. 
This rate is limited by the hardware, in particular the camera integration time of 120 ms and the time for the mechanical stage and autofocus to complete a move of about 280 ms.
The revisiting of the flakes in multiple magnifications takes the least amount of time and is negligible when compared to the time of the previous steps.

\section{Discussion}

We have presented and evaluated a newly developed setup for the automatic detection and classification of 2D material flakes.
The focus of the implementation has been on its reliability and speed, which allow its easy adaption to different laboratory settings.

The performance of our model has been evaluated in terms of the IoU, precision, and recall on a pixel and an instance level for various materials and thicknesses. 
The (pixel-wise determined) recall has been shown to be as high or surpassing those reported for implementations using neural networks \cite{Han2020Jul, Yang2022Aug, Siao2021}. 
While the high average recall demonstrates that essentially all flakes are found, the detection algorithm still wrongly detects contamination or shadows, in particular for materials of low optical contrast. 
In particular, the detection of monolayers of hBN so far remains unreliable as a result of the overwhelming number of false positives from tape residue.  
However, it might be possible by using different tapes which leave less residue, by different types of substrates, such as silicon nitride \cite{Hattori2022}, which increases the contrast.
Furthermore, the proposed pipeline allows for easy extension of the current model by either adding different post-processing steps, such as Markov Random Fields, or adding models that are specialized for different materials, such as models that utilize PL imaging for TMDs, to improve the effectiveness even further.

We have furthermore demonstrated that the algorithm can be trained on a few example images only, which allows one to add new materials relatively quickly.  
As the detection is based on a physical property of the materials, i.e. their color contrast, training could in the future even be fully replaced by simulations (zero-shot learning). 

The archiving of large amounts of standardized images of detected flakes using our approach (currently more than 65,000 flakes) furthermore has the potential to allow some degree of macro-analysis. 
It could, for example, be used to analyze properties of the respective crystals, such as preferred crystal-axis breaking \cite{Han2020Jul}. 
In addition, it could be used to improve the yield or cleanliness of common exfoliation methods through a statistical analysis of the data. 

In our laboratory, the method has been widely adapted, with more than $10^6$ images taken within 24 months by over 30 individual researchers and supporting several projects \cite{Wirth2022, Tebbe2023, Ouaj2023}. 
Our implementation not only speeds up sample fabrication for individuals, but has proven to have synergies in a group working on 2D materials, as exfoliated materials can be easily shared between individuals. 
    
\section{Acknowledgements}
This project has received funding from the European Union’s Horizon 2020 research and innovation program under grant agreement No. 881603 (Graphene Flagship) and from the European Research Council (ERC) under grant agreement No. 820254, the Deutsche Forschungsgemeinschaft (DFG, German Research Foundation) under Germany’s Excellence Strategy - Cluster of Excellence Matter and Light for Quantum Computing (ML4Q) EXC 2004/1 - 390534769, the FLAG-ERA grant TATTOOS, by the Deutsche Forschungsgemeinschaft (DFG, German Research Foundation) - 437214324. 
K.W. and T.T. acknowledge support from the JSPS KAKENHI (Grant Numbers 20H00354, 21H05233 and 23H02052) and World Premier International Research Center Initiative (WPI), MEXT, Japan.
 
\section{Data availability}
The data supporting the findings can be accessed at 10.5281/zenodo.8042835.
The code for the detection algorithm is accessible on Github \cite{code}.
A demo website showcasing the database and web interface can be accessed at \cite{website}.

\bibliography{GMM_bib}

\end{document}